\documentclass[amsmath,amssymb,aps,pra,10pt,superscriptaddress]{revtex4-2}

\usepackage{graphicx}
\usepackage{dcolumn}
\usepackage{bm}
\usepackage{footmisc}
\usepackage{lmodern}

\newcommand{\ket}[1]{\lvert#1\rangle}
\newcommand{\braket}[1]{\langle#1\rangle}
\def \a {\hat{a}}

\begin{document}

\preprint{APS/123-QED}

\title{$N$-Way Frequency Beamsplitter for Quantum Photonics}

\author{Richard Oliver}
\affiliation{Department of Applied Physics and Applied Mathematics, Columbia University, 500 W 120th St, New York, NY, 10027, US}
\author{Miri Blau}
\affiliation{Department of Applied Physics and Applied Mathematics, Columbia University, 500 W 120th St, New York, NY, 10027, US}
\author{Chaitali Joshi}
\altaffiliation{Present affiliation: Google Quantum AI, Santa Barbara, 93111, US.}
\affiliation{Department of Applied Physics and Applied Mathematics, Columbia University, 500 W 120th St, New York, NY, 10027, US}
\author{Xingchen Ji}
\altaffiliation{Present affiliation: John Hopcroft Center for Computer Science, School of Electronic Information and Electrical Engineering, Shanghai Jiao Tong University, Shanghai, 200240, China;
State Key Lab of Advanced Optical Communication Systems and Networks, Department of Electronic Engineering, Shanghai Jiao Tong University, Shanghai, 200240, China}
\affiliation{Department of Electrical Engineering, Columbia University, 500 W 120th St, New York, NY, 10027, US}
\author{\\Ricardo Guti\'errez-J\'auregui}
\affiliation{Department of Physics, Columbia University, 538 W 120th St, New York, NY, 10027, US}
\affiliation{Departamento de F\'isica Cu\'antica y Fot\'onica, Instituto de F\'isica, Universidad Nacional Aut\'onoma de M\'exico, Ciudad de M\'exico, 04510, M\'exico}
\author{Ana Asenjo-Garcia}
\affiliation{Department of Physics, Columbia University, 538 W 120th St, New York, NY, 10027, US}
\author{Michal Lipson}
\affiliation{Department of Applied Physics and Applied Mathematics, Columbia University, 500 W 120th St, New York, NY, 10027, US}
\affiliation{Department of Electrical Engineering, Columbia University, 500 W 120th St, New York, NY, 10027, US}
\author{Alexander L. Gaeta}
\email{Corresponding author: alg2207@columbia.edu}
\affiliation{Department of Applied Physics and Applied Mathematics, Columbia University, 500 W 120th St, New York, NY, 10027, US}
\affiliation{Department of Electrical Engineering, Columbia University, 500 W 120th St, New York, NY, 10027, US}

\date{\today}

\begin{abstract}
Optical networks are the leading platform for the transfer of information due to their low loss and ability to scale to many information channels using optical frequency modes. To fully leverage the quantum properties of light in this platform, it is desired to manipulate higher-dimensional superpositions by orchestrating linear, beamsplitter-type interactions between several channels simultaneously. We propose a method of achieving simultaneous, all-to-all coupling between \textit{N} optical frequency modes via \textit{N}-way Bragg-scattering four-wave mixing. By exploiting the frequency degree of freedom, additional modes can be multiplexed in an interaction medium of fixed volume and loss, avoiding the introduction of excess noise. We generalize the theory of the frequency-encoded two-mode interaction to \textit{N} modes under this four-wave mixing approach and experimentally verify the quantum nature of this scheme by demonstrating three-way multiphoton interference. The two input photons are shared among three frequency modes and display interference differing from that of two classical (coherent-state) inputs. These results show the potential of our approach for the scalability of photonic quantum information processing to general \textit{N}-mode systems in the frequency domain.
\end{abstract}

\maketitle

\section{Introduction}

Central to the viability of all quantum information platforms is the ability to scale the number information channels and their interactions without compromising performance. Regardless of physical implementation, scalability is critical for a wide variety of applications, including demonstrations of computational quantum advantage \cite{arute2019,pan2019,pan2020,madsen2022}, fault tolerance \cite{fowler2012, acharya2023}, quantum metrology and distributed sensing \cite{giovannetti2006,giovannetti2011,ge2018,zhuang2018}, and quantum communication \cite{tittel2000,cerf2002,groblacher2006,sheridan2010,mirhosseini2015,ali-khan2007,islam2017,lee2019,hu2018,cozzolino2019}.

In recent years, quantum photonic experiments have made progress towards scalability using both discrete \cite{pan2019,bartolucci2023} and continuous variables \cite{zhuang2018,larsen2019,pan2020,yokoyama2013,madsen2022}, positioning photonic systems as a promising platform for quantum information processing. Light enjoys numerous advantages as an information carrier including low decoherence due to limited interactions with its environment. Nevertheless, nonlinear interactions between photons tend to be far weaker than linear ones, a fact which leans proposals for photonic quantum computing towards measurement-based models in which measurements effect nonlinear interactions \cite{knill2001, raussendorf2001}.  Whatever the computational viability of optical qubits, optical quantum networks are an ideal way of providing interconnections due to the low loss and dispersion of the telecom network, and optical interferometers will continue to serve as natural platforms for quantum metrology as they have throughout the history of the field. All of these applications make use of linear coupling between optical modes, conventionally achieved via sequential two-port beamsplitters. However, since such interactions are limited to coupling two modes at a time (e.g. polarization or spatial modes), additional losses are incurred with each added interaction, impeding scalability.

Interference between two frequency modes, in particular frequency-domain Hong-Ou-Mandel interference \cite{HOM1987}, has been proposed and experimentally established using both four-wave mixing and electro-optic approaches \cite{raymer2010,kobayashi2016,imany2018,lu2018,joshi2020}. While frequency-domain interference beyond two modes has been treated both theoretically and experimentally using four-wave mixing in the classical regime \cite{bell2017,wang2020}, as well as using electro-optic modulation in both classical and quantum regimes with single or parallel qudits \cite{yuan2016,lu2018,hu2020,hu2021,wang2020,buddhiraju2021,yuan2021,lu2022a,lu2022b}, frequency-domain interference of correlated photon pairs in more than two modes has not been shown.

\begin{figure}[ht]%
\centering
\includegraphics[width=\textwidth]{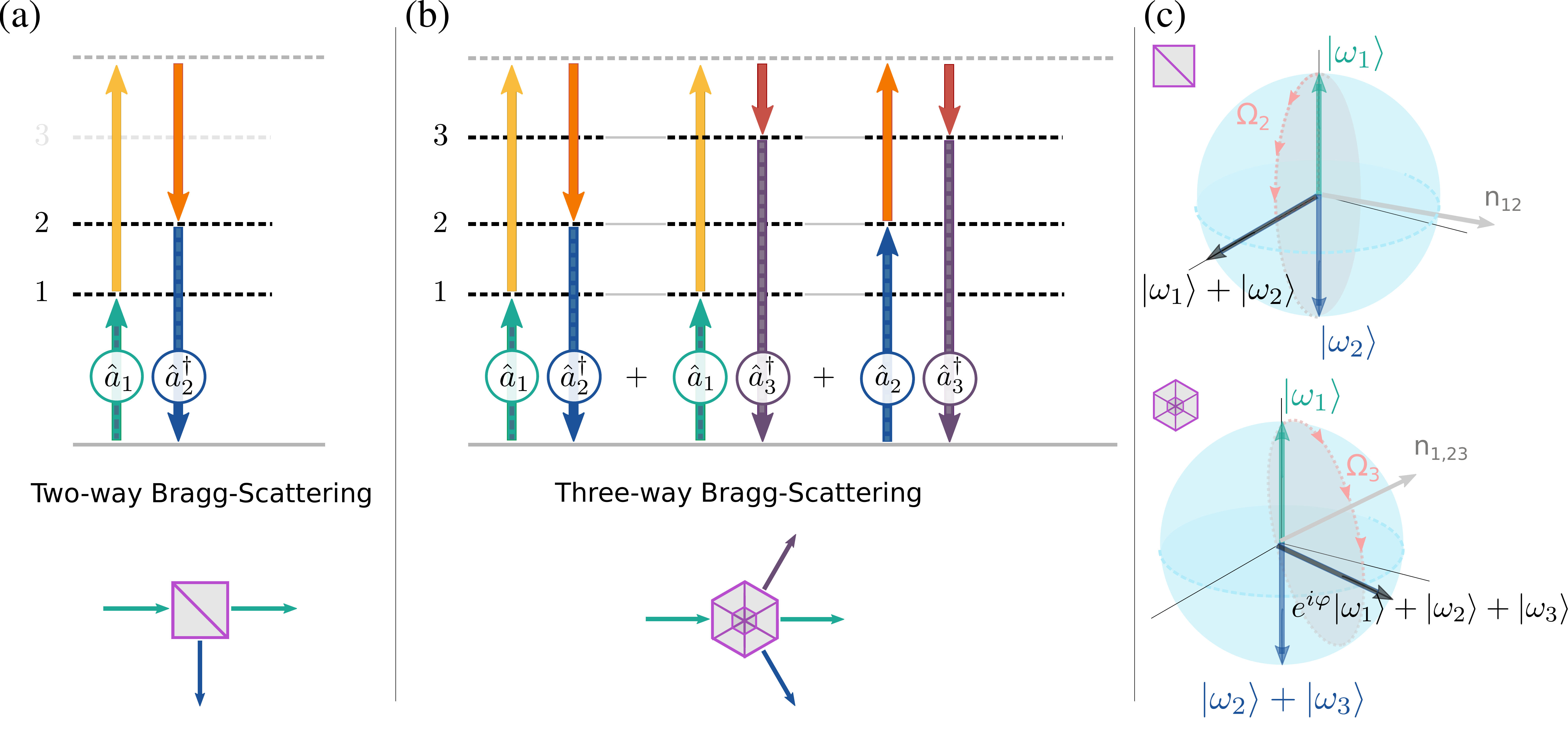}
\caption{Energy-level diagrams for the two-way (a) and three-way (b) Bragg-scattering four-wave mixing process (BS-FWM), where a photon (dashed arrows) is transferred from one mode to another with the aid of two pump beams (solid arrows) satisfying phase-matching conditions. This nonlinear, pairwise process is extended in (b) to occur simultaneously through three different channels. The process is coherent, allowing for excitations to be exchanged between the modes and can be understood in analogy to a frequency-domain beam-splitter (bottom panels). For simplicity the reverse creation/destruction processes are not shown. (c) When the modes share a single excitation, the coherent transfer is represented by the Bloch sphere, where the north pole represents the initial mode and the south pole represents either the second mode (two-way) or a superposition of the second and third modes (three-way).
}
\label{energy}
\end{figure}

We move beyond the two-mode limit by implementing N-way Bragg-scattering four-wave mixing (BS-FWM). This process coherently \cite{clemmen2016} interacts optical modes of different frequencies and can be viewed as a beamsplitter in the frequency domain. Whereas conventional beamsplitters such as polarizing beamsplitters are limited to interfering two modes at a time, N-way BS-FWM couples all pairs among $N$ modes selected from a continuum and multiplexed together, which enables control over a full bosonic $N$-level system (Fig. \ref{energy}). We theoretically articulate the extension of the basic $2\times2$ Hamiltonian to an $N\times N$ one, demonstrating it experimentally for $N=3$. Not only can this method extend the number of interacting modes and induce an all-to-all interaction, but it also has the advantage of allowing frequency conversion over hundreds of GHz, making it compatible with the frequency spacings of photons generated by existing small-footprint ($\sim$100 $\mu$m-radius) integrated microresonators. Optical multiport interferometers have been extensively studied \cite{zukowski1997,tichy2010} with experimental realizations focusing particularly on spatial \cite{weihs1996,spagnolo2012,meany2012,crespi2013,metcalf2013,spagnolo2013,menssen2017} and temporal \cite{brecht2014,brecht2015,serino2023,folge2024} modes. Unlike the $\chi^{(2)}$ proposal for generating a frequency beamsplitter in \cite{folge2024}, in our approach the sets of input and output frequency modes are identical. This ensures symmetric optics constraints on the input state as on the output and readily allows for incorporation within existing optical networks designed for fixed frequency grids within a single optical band.

Bragg-scattering four-wave mixing (BS-FWM) is a noiseless, nonlinear process capable of performing unitary transformations between two optical frequency modes interacting in a third-order ($\chi^{(3)}$) nonlinear medium (Fig. \ref{energy}) \cite{marhic1996,mckinstrie2005,mcguinness2010,raymer2010,clark2013,donvalkar2014,agha2012,li2016,bell2016,zhao2022}. Since a nonlinear fiber is capable of extremely low propagation losses ($<1\%$ for the 100-m length of fiber in this experiment), deviations from unitary, noiseless operation are minute. This frequency-conversion process is depicted in Fig. \ref{energy}a, where two high-intensity pump fields (solid yellow and orange) induce the coherent transfer of population between two frequency modes, traditionally termed the signal and idler (dashed green and blue). In the frequency domain, this process is identical to a beamsplitter, for which the input and output ports correspond to frequency channels. In the case of a single-photon excitation, the signal/idler frequency modes constitute a two-level system, and the four-wave-mixing interaction is equivalent to a rotation on the Bloch sphere in which the poles represent the two modes while all other points correspond to coherent superpositions \cite{clemmen2016}. An efficient population transfer must satisfy energy and momentum conservation, known as phase-matching. By simultaneously phase-matching multiple processes we can realize a three-way frequency beamsplitter, represented graphically in Fig. \ref{energy}b. Alternatively, this may be viewed as a transformation on a three-level system, without adding noise or loss.

\section{Theory}

The theory of BS-FWM among $N$ modes can be developed from the two-mode case. The dynamics of Bragg scattering between two frequency modes is given by the spatial propagation Hamiltonian,
$\hat{H}_{N=2}=2\gamma A_1 A_2^* \a_1 \a_2^{\dag} + H.a.$, along with the spatial Heisenberg equations \cite{mckinstrie2005} for a given operator $\hat{O}$, $d\hat{O}/dz=i[\hat{O},\hat{H}]$.
In this Hamiltonian, $A_i$ represents the complex amplitude of the $i^{th}$ pump field (units of W$^{1/2}$), $\gamma$ is the effective nonlinearity ($\text{W}^{-1}\text{m}^{-1}$), and $\a_i(z)$ is the bosonic operator annihilating the $i^{th}$ frequency mode at the position $z$ as it propagates along the fiber. The operators $\a_i$ obey equal-position commutation relations $[\a_i(z),\a^{\dag}_j(z)]=\delta_{ij}$. The Heisenberg-picture dynamics for the above Hamiltonian induce the following relationship between the input and output fields, where $L$ is the length of the nonlinear medium:

\begin{equation}
    \begin{bmatrix}
        \a_1(L)\\
        \a_2(L)
    \end{bmatrix}
    =\exp\left\{iL
    \begin{bmatrix}
        0 & 2\gamma A_1^* A_2 \\
        2\gamma A_1 A_2^* & 0
    \end{bmatrix}
    \right\}
    \begin{bmatrix}
        \a_1(0)\\
        \a_2(0)
    \end{bmatrix}.
    \label{eqn:2dimdyn}
\end{equation}

\noindent The above process describes a matrix transformation of the input signal and idler
fields, $\a_i$ and is equivalent to a beamsplitter with input ports corresponding to the two input frequencies \cite{raymer2010,mcguinness2010}. Since the pumps are in a high-intensity coherent state, quantum features from the pumps are negligible so that the pump fields are represented as classical numbers.

The two-input case can be extended to that of $N$ fields interacting via BS-FWM through the following Hamiltonian:

\begin{equation}
    \hat{H}_N=\sum_{i\ne j}2\gamma A_i A_j^* \a_i\a_j^{\dag}.
\end{equation}
\noindent Note that although the effective nonlinearity $\gamma$ is wavelength-dependent, in this equation it corresponds to the carrier wavelength and is therefore treated as a constant of the fiber. As in the case of two modes, the Heisenberg equations of motion yield an $SU(N)$ transformation between the $N$ interacting fields such that,

\begin{equation}
    \begin{bmatrix}
        \a_1(L)\\
        \vdots \\
        \a_N(L)
    \end{bmatrix}
    =\exp\left\{i L
    \begin{bmatrix}
        0 & & 2\gamma A_i^* A_j \\
        & \ddots &\\
        2\gamma A_i A_j^* & & 0
    \end{bmatrix}
    \right\}
    \begin{bmatrix}
        \a_1(0)\\
        \vdots\\
        \a_N(0)
    \end{bmatrix}.
    \label{eqn:Npump}
\end{equation}

\noindent The inclusion of phase-matching introduces real diagonal elements in the above matrix (see Supplementary Material), which may be neglected in our experiment by suitable choice of wavelengths. Since the phase-matching condition and interaction strength of one pair (say between modes one and three) are determined by the phase-matching condition of the other two pairs (one-two and two-three), there is a limitation to the set of available transformations. This can also be seen from the observation that the number of degrees of freedom in the transformation matrix of Eq. \ref{eqn:Npump} is $\mathcal{O}(N)$, whereas an arbitrary $SU(N)$ transformation is parametrized by $\mathcal{O}(N^2)$ numbers.

For the case of $N=3$ pumps, assuming negligible phase mismatch and equal pump powers $\lvert A_i\rvert=P$, the transformation may be written as a function of nonlinear phase $\phi=2\gamma L P$ (see Supplementary Material):
\begin{equation}
    \begin{bmatrix}
        \a_1(L)\\
        \a_2(L)\\
        \a_3(L)
    \end{bmatrix}
    =
    \begin{bmatrix}
        p(\phi) & q(\phi) & q(\phi)\\
        q(\phi) & p(\phi) & q(\phi)\\
        q(\phi) & q(\phi) & p(\phi)
    \end{bmatrix}
    \begin{bmatrix}
        \a_1(0)\\
        \a_2(0)\\
        \a_3(0)
    \end{bmatrix},
    \label{eqn:transformation}
\end{equation}
where $q(\phi)=(e^{3i\phi}-1)/3$, $p(\phi)=q(\phi)+1$.

Whereas a polarizing beamsplitter is restricted to two input ports, we demonstrate that a frequency-domain beamsplitter may couple more than two modes within the same nonlinear medium. A frequency-domain beamsplitter therefore scales favorably with respect to loss and complexity in comparison with concatenating spatial-mode beamsplitters for $N$-mode operation since additional ports can be multiplexed in the same waveguide, keeping propagation loss fixed in a single device with the same physical size \cite{joshi2018}.

To demonstrate the all-to-all interaction, we consider both attenuated coherent states as well as photons as inputs. For a single-frequency input in mode one, the populations in the three frequency modes are given by $\braket{\a_1^{\dagger}\a_1}=\lvert p(\phi)\lvert ^2$, $\braket{\a_2^{\dagger}\a_2}=\braket{\a_3^{\dagger}\a_3}=\lvert q(\phi)\lvert ^2$. Experimentally, these correspond to single-click detection events (``singles rates"). These expressions hold for both coherent states and single photons and represent tunable transmission coefficients of a three-way beamsplitter.

Extending to dual-frequency inputs, in which modes one and three are excited, we now compute both the first- and second-order correlation functions corresponding to either two attenuated coherent states or a photon pair. The analytical results for both correlation functions (singles and coincidences) are tabulated in Table \ref{tbl:theory} and assume equal intensities of each input channel (in addition to the assumptions behind Eq. \ref{eqn:transformation}). The first-order correlation functions (singles counts) are identical; only in the second-order correlation functions or coincidences are non-classical statistics revealed. The coherent states yield uncorrelated or accidental coincidences given by the product of their singles counts, whereas correlated photon pairs show genuine two-photon interference, an extension of the Hong-Ou-Mandel effect.

\begin{table}[ht]
\centering
\caption{Theoretical detection statistics of the dual-frequency state after Bragg scattering, corresponding to first- and second-order correlation functions. Note that by the symmetry of the inputs $\braket{\a_1^{\dagger}\a_1}=\braket{\a_3^{\dagger}\a_3}$ and $\braket{\a_1^{\dagger}\a_2^{\dagger}\a_2\a_1}=\braket{\a_2^{\dagger}\a_3^{\dagger}\a_3\a_2}$.}
\label{tbl:theory}
\begin{tabular}{l|ll}
\toprule
 & Dual-Freq. Coherent State  & Photon Pairs\\
\hline
Singles & $\braket{\a_1^{\dagger}\a_1}=1-\lvert q(\phi)\lvert^2$ & $\braket{\a_1^{\dagger}\a_1}=1-\lvert q(\phi)\lvert^2$\\
& $\braket{\a_2^{\dagger}\a_2}=2 \lvert q(\phi)\lvert ^2$ & $\braket{\a_2^{\dagger}\a_2}=2 \lvert q(\phi)\lvert ^2$\\
Coincidences & $\braket{\a_i^{\dagger}\a_j^{\dagger}\a_j\a_i}=\braket{\a_i^{\dagger}\a_i}\braket{\a_j^{\dagger}\a_j}$ & $\braket{\a_1^{\dagger}\a_2^{\dagger}\a_2\a_1}=\lvert p(\phi)q(\phi)+q^2(\phi)\lvert^2$\\
& & $\braket{\a_1^{\dagger}\a_3^{\dagger}\a_3\a_1}=\lvert p^2(\phi)+q^2(\phi)\lvert^2$\\
\botrule
\end{tabular}
\end{table}

\section{Results}

\begin{figure}[ht]
\centering
\includegraphics[width=\textwidth]{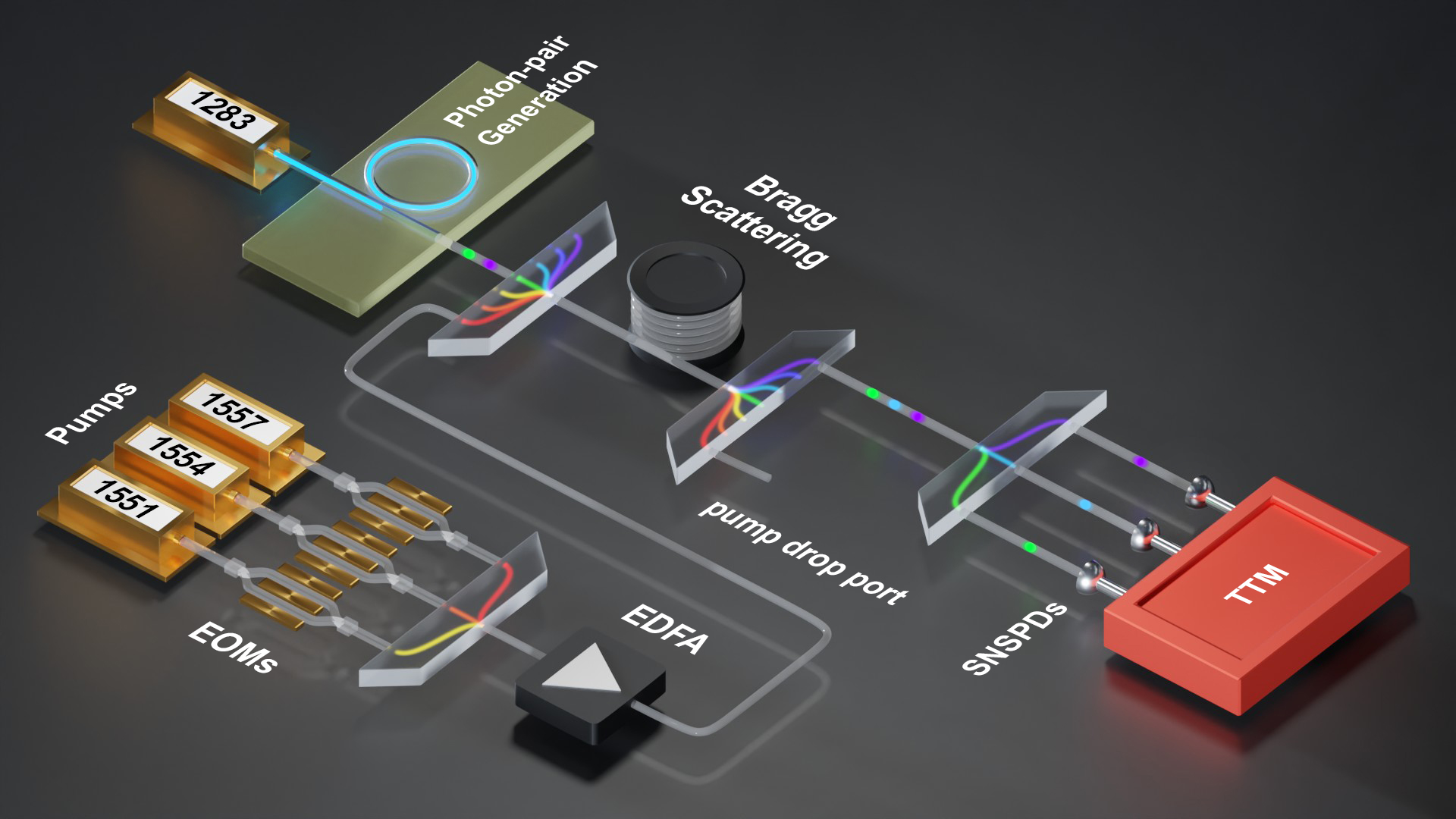}
\caption{Three-Way Bragg-Scattering Four-Wave Mixing. Two photons are sent into two of the three frequency ports of the nonlinear fiber. Before the fiber they are wavelength-multiplexed with the three Bragg pumps, shown at bottom left being modulated and then amplified to generate high-peak-power pulses. Following the fiber the pumps are filtered out of the system and the photons generated at the three output channels are filtered and sent to their respective detectors. EOM, electro-optic modulator; EDFA, erbium-doped fiber amplifier; SNSPD, superconducting nanowire single-photon detector; TTM, time-tagging module.}
\label{setup}
\end{figure}

\begin{figure}[ht]%
\centering
\includegraphics[width=\textwidth]{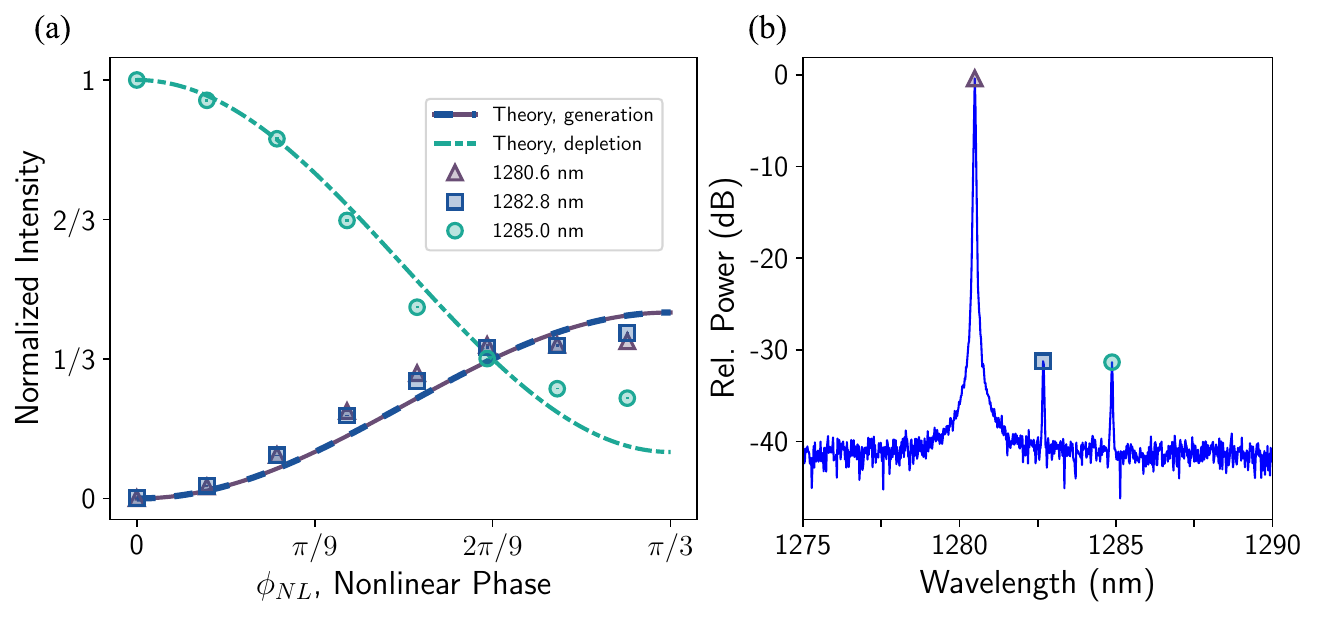}
\caption{(a) Normalized rates of single-click events as a function of the nonlinear phase $\phi_{NL}$. The classical input (here an attenuated coherent state at 1285.0 nm) is converted to the other two signals (at 1280.6 nm, purple, and 1282.8 nm, blue). (b) Optical spectrum of the output fields after a nonlinear interaction such that the three frequency modes share nearly equal power, corresponding to $\phi_{NL}=2\pi/9$. The difference in average power level between the continuous-wave input (here 1280.6 nm) and the outputs at the converted frequencies (1282.8 nm, 1285.0 nm) corresponds to the duty cycle of the pump pulses and therefore that of the newly generated output pulses. The all-to-all interaction is controlled by using phase-matching to confine the coupling to the three frequencies shown; coupling to external frequency modes is negligible.}
\label{sgls_one_freq}
\end{figure}

\begin{figure}[ht]%
\centering
\includegraphics[width=0.5\textwidth]{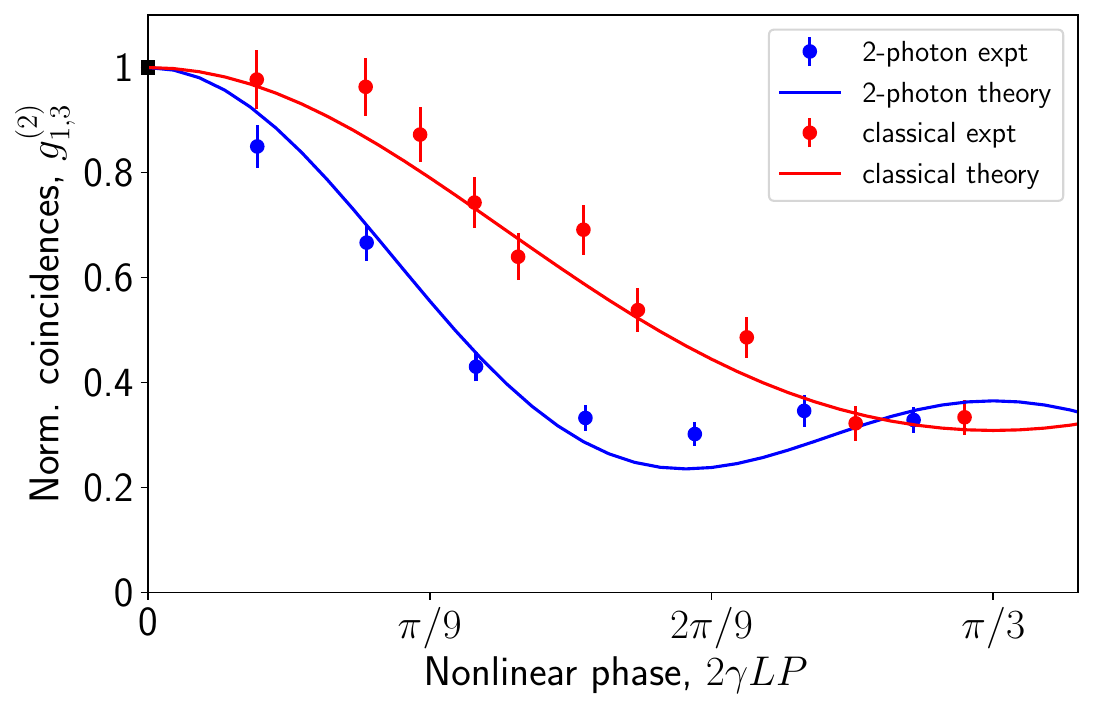}
\caption{Measured coincidences (integrated $g^{(2)}$ normalized to rates at zero nonlinear phase) for photon pairs (blue) versus two classical, attenuated coherent states (red) after N-way BSFWM.}
\label{coinc_two_freq}
\end{figure}

We experimentally demonstrate a three-way interaction via Bragg-scattering FWM to realize all-to-all coupling in the three-mode system. Fig. \ref{setup} shows our experimental setup with three frequency modes in the O-band (1260-1360 nm). Our correlated photon pairs are generated through spontaneous four-wave mixing (SFWM) in a SiN microresonator. The microresonator is pumped with a CW laser at 1282.8 nm, and the generated photons are combined with the three strong classical pump fields located in the C band (1530-1565 nm) and sent to a dispersion-shifted fiber (Corning Vistacor) for a nearly perfectly phase-matched nonlinear interaction (see Fig. \ref{setup}). The powers of the classical pumps are scanned to obtain the nonlinear-phase-dependent interference pattern (Figs. \ref{sgls_one_freq}, \ref{coinc_two_freq}). After the nonlinear interaction, the three frequency arms are separated, filtered, and detected with superconducting nanowire single photon detectors synchronized with a time tagging module for coincidence counting. 

We begin by exciting our system in the classical regime by injecting a single-wavelength weak coherent state and measuring the energy exchange between the three frequency modes while varying the nonlinear phase $\phi_{NL}$. As the nonlinear phase increases, the input signal is converted to the other wavelength channels (see Fig. \ref{sgls_one_freq}a). For a nonlinear phase of $\phi_{NL}=2\pi/9$, power in the input signal is shared evenly between all three frequency modes, equivalent to a frequency-domain tritter \cite{lu2018}. The measured spectrum confirms that Bragg scattering is limited to the three channels of interest. Further, the spectrum confirms that power from the input signal is symmetrically distributed between the other channels. Note that the difference in spectral peak heights is due to the pulsing of the BS-FWM process, so that the generated idlers are pulses with a duty cycle of 26.2 dB. The additional contribution to peak height is the relative generation rate factor (compared to CW input) of approximately one third. The spectrum is thus essential in confirming these properties of our three-way nonlinear process.

Observable differences arise between the statistics of coherent states and single-photon states. Whereas the above first-order interferences do not discern classical weak coherent states from photon pairs, the second-order correlation measurements do distinguish these inputs, which we illustrate by performing integrated $g^{(2)}$ measurements, shown in Fig. \ref{coinc_two_freq} for each type of input state. The coincidence rates of the weak coherent state are given by the square of the singles rates, as expected for a correlation of two independent variables. Meanwhile, the coincidence rates for the photon-pair input notably decrease faster as a function of nonlinear phase. This feature may be interpreted as an extension to the three-mode system of Hong-Ou-Mandel interference, in which coincidences vanish at the nonlinear phase corresponding to a 50:50 beamsplitter \cite{joshi2020}. The interference curve is lifted further by the nonzero multi-photon component simultaneously with the asymmetric loss on each frequency channel following pair generation and prior to interference, as predicted by theory in the Supplementary Material. After carefully accounting for these effects, we observe excellent agreement with theory.

\section{Discussion}

We propose an all-to-all coherent photonic interaction between $N$ frequency modes and experimentally demonstrate the system for three modes. The coupling is low-noise and allows for coherent transfer of population between the modes. The available unitary transformations in our scheme are dictated by phase-matching considerations. Phase-matching is a momentum conservation phenomenon, and as such, requires careful dispersion engineering to tailor the desired transformations. In our case, the phase-matching conditions are constrained by higher-order dispersion. In the case we realize here, we phase-match all three pairs of frequency modes.

To practically scale beyond three modes, it is possible to take advantage of the higher-order phase-matching such that each pair of modes may be coupled by a unique pair of pumps. This places only a classical overhead on the system, to be expected from the fundamental requirements for an $N\times N$ transformation. Most importantly, adding to the number of pumps allows for scaling the number of beamsplitters without introducing additional losses. Further, owing to the fact that Bragg-scattering can couple optical modes separated by 100s of GHz, our scheme is compatible with typical resonance spacings of integrated microresonator-based quantum sources, as shown in this experiment. As BSFWM has already been demonstrated in an integrated platform (including suitability for use with quantum states) \cite{agha2012,li2016,zhao2022}, we envision full integration of quantum state generation and manipulation using our scheme.

Our measurements represent the first demonstration of two-photon interference across more than two modes in the frequency domain using N-way BS-FWM. We show theoretically how interfering $N$ modes is possible using the same number of pump modes, opening the door to a demonstration of quantum advantage via frequency-domain boson sampling. This work therefore represents a pathway towards scalability of photonic quantum information processing in the frequency domain.
\\
\\
\indent This research was supported in part by the National Science Foundation (OMA-1936345, PHY-2110615) and the Department of Energy, Office of Science, National Quantum Information Science Research Centers, Co-design Center for Quantum Advantage (C$^2$QA).

\bibliography{bib}

\end{document}


\title{Supplementary Material}

\author{Richard Oliver$^1$, Miri Blau$^1$, Chaitali Joshi$^1$, Xingchen Ji$^2$,\\
Ricardo Guti\'errez-J\'auregui$^{3,4}$, Ana Asenjo-Garcia$^3$, Michal Lipson$^{1,2}$, Alexander L. Gaeta$^{1,2}$}
\affil{
$^1$Department of Applied Physics and Applied Mathematics, Columbia University, 500 W 120th St, New York, NY, 10027, US\\
$^2$Department of Electrical Engineering, Columbia University, 500 W 120th St, New York, NY, 10027, US\\
$^3$Department of Physics, Columbia University, 538 W 120th St, New York, NY, 10027, US\\
$^4$Departamento de F\'isica Cu\'antica y Fot\'onica, Instituto de F\'isica, Universidad Nacional Aut\'onoma de M\'exico, Ciudad de M\'exico, 04510, M\'exico
}

\maketitle

\section{Classical Derivation of N-Pump BS-FWM}

We will consider an electromagnetic wave propagating in a nonlinear fiber. In particular we will model the dynamics of the slowly-varying envelope $A(z,t)$ of the electric field (assuming a single polarization mode)
\begin{equation}
    E(\vec{r},t)=\frac{1}{2}F(x,y)A(z,t)e^{i(\beta_0 z-\omega_0 t)}+c.c.,
\end{equation}
with spatial mode profile given by $F(x,y)$ and defined with respect to the fast carrier-signal wavevector $\beta_0$ and frequency $\omega_0$. The envelope $A(z,t)$ is well-described classically by the nonlinear Schr\"odinger equation \cite{agrawal2019}:
\begin{equation}
    \partial_z A(z,t) = \left(i\sum_{m=1}^{\infty}\frac{\beta_m}{m!}(   i\partial_t)^m+i\gamma|A(z,t)|^2\right)A(z,t),
    \label{eqn:NLSE}
\end{equation}
written in terms of the effective nonlinearity $\gamma$ and the coeficients of the fiber dispersion relation:
\begin{equation}
    \beta(\omega)=\sum_m \tfrac{\beta_m}{m!}(\omega-\omega_0)^m,
\end{equation}
where $\beta_m:=d^m\beta/d\omega^m|_{\omega_0}$.

The field envelope may be expanded generally into its frequency components:
\begin{equation}
    A(z,t)=\sum_j A_j(z) e^{i(\delta\beta_j z-\delta\omega_j t)},
    \label{eqn:sve_decomp}
\end{equation}
where $\delta\beta_j:=\beta(\omega_j)-\beta_0$ and $\delta\omega_j:=\omega_j-\omega_0$ are the wavevector and frequency offsets from those of the carrier. While these frequency components are independent in linear media, they are coupled by optical nonlinearity. This is evident upon inserting Eq. (\ref{eqn:sve_decomp}) into Eq. (\ref{eqn:NLSE}) and matching up Fourier components, which yields:
\begin{equation}
    \partial_z A_n(z)=i\gamma\sum_{\left\{\substack{k,l, m:\\ \omega_k+\omega_l=\\ \omega_m+\omega_n}\right\}}e^{i\Delta \beta^{kl}_{mn} z} A_k(z)A_l(z)A_m^*(z).
    \label{eqn:fwm}
\end{equation}
In Eq. (\ref{eqn:fwm}), we introduce the wavevector mismatch:
\begin{equation}
    \Delta \beta^{kl}_{mn}:=\beta(\omega_k)+\beta(\omega_l)-\beta(\omega_m)-\beta(\omega_n).
\end{equation}
The conditions $\omega_k+\omega_l=\omega_m+\omega_n$ and $\Delta\beta^{kl}_{mn}=0$ may be interpreted respectively as energy and momentum conservation for a process in which particles $k$ and $l$ collide and reemerge as particles $m$ and $n$. These conditions constrain the nonlinear interactions contained by the sum in Eq. (\ref{eqn:fwm}). Satisfying these constraints is known as phasematching.

The dominant nonlinear processes are those which satisfy momentum conservation, allowing efficient interaction throughout propagation. This may be seen by integrating Eq. 
(\ref{eqn:fwm}) with respect to $z$ to obtain the $n^{th}$ field after a certain propagation distance $L$:
\begin{equation}
    A_n(L)=i\gamma\int_0^L dz\sum_{\left\{\substack{k,l, m:\\ \omega_k+\omega_l=\\ \omega_m+\omega_n}\right\}}e^{i\Delta \beta^{kl}_{mn} z} A_k(z)A_l(z)A_m^*(z).
\end{equation}
Those terms on the RHS for which $\Delta \beta^{kl}_{mn}$ is large in magnitude will carry a rapidly oscillating phasor $e^{i\Delta \beta^{kl}_{mn} z}$; if it oscillates faster than the fields $A_j$ are changing, then the phasor will interfere destructively over the course of the propagation and its contribution to the integral will be negligible. Therefore, only terms for which $\Delta\beta^{kl}_{mn}\ll \sim 1/L$ will be important; all others may be dropped from the sum in the RHS of Eq. (\ref{eqn:fwm}). Long propagation lengths typical for fibers are therefore advantageous in constraining the many possible nonlinear processes to those which are desired through phasematching.

So far, we have presented a general description of light propagating in a one-dimensional, dispersive, lossless, third-order nonlinear medium. We now consider our particular setup where the total field consists of high-intensity 
``pump" fields and low-intensity ``weak" fields:
\begin{equation}
    A(z,t)=\underbrace{\sum_{p\in P} A_p(z) e^{i(\delta\beta_p z-\delta\omega_p t)}}_{\textrm{pumps}} + \underbrace{\sum_{s\in S} b_s(z) e^{i(\delta\beta_s z-\delta\omega_s t)}}_{\textrm{weak fields}}.
    \label{eqn:Adef}
\end{equation}
Here, $P$ and $S$ denote the set of pump and weak fields, respectively. (The weak fields will later become the signal and idler fields in two-pump Bragg scattering.) First, let us recall the case of Bragg scattering with two pumps \cite{mckinstrie2005,farsi2015,joshithesis2020}; in order for this process to be phasematched it is sufficient to situate the pumps and the weak signal/idler fields approximately symmetrically about the zero-group velocity dispersion frequency $\omega_{ZGVD}$, defined by $\beta_2(\omega_{ZGVD}):=0$. Since the nonlinear medium couples different weak fields in pairs, the case of $N$ pumps considered in this work requires that each pair of pumps be displaced symmetrically from the corresponding pair of weak fields about $\omega_{ZGVD}$ \cite{joshithesis2020}.

\begin{figure}[ht]
\centering
\includegraphics[width=0.8\textwidth]{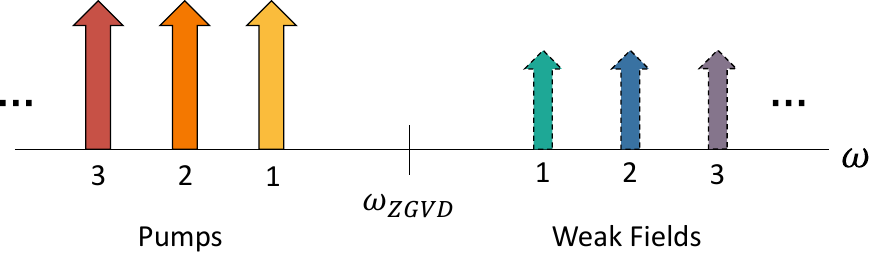}
\caption{Phasematching requires symmetric locations of the pumps and weak signal/idler frequencies about the zero-group velocity dispersion frequency, $\omega_{ZGVD}$.}
\label{phasematching_scheme}
\end{figure}

Motivated by this phasematching scheme, we consider an array of $N$ pumps mixing $N$ weak fields, shown in Fig. \ref{phasematching_scheme} along with the labeling convention. We start by finding the steady-state solution for the pump fields. Starting from Eq. (\ref{eqn:fwm}) we include only phasematched terms under the approximation that contributions due to the weak fields are negligible compared to those of the pumps themselves and cannot appreciably remove energy from the pump modes. This results in:
\begin{equation}
    \partial_z A_n(z)=i\gamma\big(\overbrace{|A_n(z)|^2}^{\textrm{self-phase}} + \sum_{p\ne n} \overbrace{2|A_p(z)|^2}^{\textrm{cross-phase}}\big)A_n(z).
    \label{eqn:pumpdynamics}
\end{equation}
From Eq. (\ref{eqn:pumpdynamics}) it follows that
\begin{equation}
    \partial_z |A_n(z)|^2=A_n^*(z)\partial_z A_n(z)+c.c.=0\,;
    \label{eqn:intensity_dynamics}
\end{equation}
this yields $|A_n(z)|=|A_n(0)|$. The pump fields therefore only change by a phase. This is a consequence of the above approximation in which weak field contributions to pumps dynamics were neglected. Because this implies that the pump fields cannot lose energy it is known as the undepleted-pump approximation \cite{boyd}. This approximation is particularly reliable for BS-FWM since only a relatively small and finite amount of power is taken from the pumps in the conversion from signal to idler. This power is a fraction at most $\sim \frac{P_{sig} (\omega_{idl}-\omega_{sig})}{P_{pump}\omega_{sig}}\ll1$).

Proceeding, the solution to Eq. (\ref{eqn:pumpdynamics}) is given in terms of the initial pump amplitudes:
\begin{align}
    A_n(z)&=e^{i\Gamma_n z}A_n(0),\\
    \Gamma_n&=\gamma\big(|A_n(0
    )|^2+2\sum_{p\ne n}|A_p(0)|^2\big).
\end{align}

Turning now to the weak-field dynamics, Eq. (\ref{eqn:fwm}) leads to:
\begin{equation}
    \partial_z b_n(z)=i\gamma\bigg(\sum_p\overbrace{2|A_p(z)|^2 b_n(z)}^\textrm{cross-phase}+\sum_{\left\{\substack{j,k,l:\\l\ne n,\\ \omega_j+\omega_l=\\ \omega_k+\omega_n}\right\}}\overbrace{2e^{i\Delta\beta^{jl}_{kn} z}A_j(z) A^*_k(z) b_l(z)}^\textrm{Bragg-scattering}\bigg),
    \label{eqn:weakfielddynamics}
\end{equation}
where, as with the pumps, terms higher than lowest order in the weak fields may be neglected (due to the large intensity of the pump fields). Using $\Omega_i$ and $\omega_i$ for the $i^{th}$ pump and weak-field frequency respectively, the above may be simplified considerably on rewriting the energy conservation condition: $\Omega_j+\omega_l=\Omega_k+\omega_n$. From the assumed configuration of frequencies, energy and momentum are both conserved only when $j=l$ and $k=n$. Thus we may write:
\begin{equation}
    \sum_{\left\{\substack{j,k,l:\\l\ne n,\\ \omega_j+\omega_l=\\ \omega_k+\omega_n}\right\}}\rightarrow
    \sum_{\left\{\substack{j,k,l:\\l\ne n}\right\}} \delta_{jl}\delta_{kn}.
\end{equation}
Then defining the $N$ rotating frames
\begin{align}
    \bb_n(z)&:=b_n(z) e^{i \varphi_n z},\\
    \varphi_n&:=\Delta\beta^n_1+\gamma\big(P_1-P_n-2\sum_p P_p\big),\\
    \Delta\beta^n_m&:=\beta(\Omega_n)+\beta(\omega_n)-\beta(\Omega_m)-\beta(\omega_m),
\end{align}
where $P_p=|A_p(0)|^2$ is the power of the $p^{th}$ pump, one obtains the simpler form
\begin{equation}
    \partial_z \bb_n(z)=i\Delta k_{n}\bb_n(z)+
    2i\gamma\sum_{l\ne n} A_l(0) A^*_n(0) \bb_l(z),
    \label{eqn:componentevol}
\end{equation}
where $\Delta k_n$ is the nonlinear phase mismatch:
\begin{equation}
    \Delta k_n:=\Delta\beta^n_1+\gamma(P_1-P_n).
\end{equation}
It is then evident that the weak fields undergo unitary evolution via the matrix equation
\begin{equation}
    \partial_z
    \begin{bmatrix}
        \bb_1(z)\\
        \vdots\\
        \bb_N(z)
    \end{bmatrix}
    =
    i\begin{bmatrix}
        0 & & & 2\gamma A_i^*(0) A_j(0)\\
        & \Delta k_2 & & \\
        & & \ddots & \\
        2\gamma A_i^*(0) A_j(0) & & & \Delta k_N
    \end{bmatrix}
    \begin{bmatrix}
        \bb_1(z)\\
        \vdots\\
        \bb_N(z)
    \end{bmatrix},
    \label{eqn:Npump}
\end{equation}
which reduces to the conventional Bragg-scattering transformation \cite{mckinstrie2005} for $N=2$ as expected. Note that the dependence on relative pump phases in Eq. (\ref{eqn:Npump}) can be factored out of the transformation by rewriting Eq. (\ref{eqn:componentevol}):
\begin{equation}
    \partial_z\bb_n(z)=i\sum_l\bigg[\delta_{nl}(\Delta k_{n}-2\gamma P_n)+
    2\gamma \sqrt{P_n P_l} \bigg]e^{i(\theta_l-\theta_n)}\bb_l(z),
\end{equation}
\noindent with the relative phases $\theta_i$, defined by $A_n(0)=\sqrt{P_n}e^{i\theta_n}$, inherited from the pump fields.

For simplicity, we consider the case of perfect phasematching $\Delta k_n=0$ and uniform pump powers, $P_n=P$. The propagation equation is then (factoring out the relative pump phases $\theta_n$ and grouping them with the $\bb_n$):
\begin{equation}
    \partial_z
    \begin{bmatrix}
        \bb_1(z)e^{i\theta_1}\\
        \vdots\\
        \bb_N(z)e^{i\theta_N}
    \end{bmatrix}
    =2i\gamma P
    \begin{bmatrix}
        0 & 1 & 1\\
        1 & \ddots & 1\\
        1 & 1 & 0\\
    \end{bmatrix}
    \begin{bmatrix}
        \bb_1(z)e^{i\theta_1}\\
        \vdots\\
        \bb_N(z)e^{i\theta_N}
    \end{bmatrix}.
\end{equation}
The solution\footnote{Note that for the $N$-dimensional matrix defined by $O_{ij}=1$, it follows that $O^2=NO$.} at $z=L$ is given in terms of the nonlinear phase $\phi=2\gamma L P$:
\begin{equation}
\begin{bmatrix}
    \bb_1(L)e^{i\theta_1}\\
    \vdots\\
    \bb_N(L)e^{i\theta_N}
\end{bmatrix}
=
e^{-i\phi}
\begin{bmatrix}
    p_N(\phi) & & q_N(\phi)\\
     & \ddots & \\
    q_N(\phi) & & p_N(\phi)\\
\end{bmatrix}
\begin{bmatrix}
    \bb_1(0)e^{i\theta_1}\\
    \vdots\\
    \bb_N(0)e^{i\theta_N}
\end{bmatrix},
\end{equation}
where
\begin{align}
    q_N(\phi)=\frac{e^{i N\phi}-1}{N},\\
    p_N(\phi)=q_N(\phi)+1.
\end{align}
Observe that the combined power from all $N$ pumps contributes to the total nonlinear phase; it is $N\phi$ that dictates the conversion strength.

\subsection{Incorporating Loss}

To understand the effects of propagation loss for the case of perfect phasematching with $N$ pumps of equal initial power, we modify the dynamics of the pumps and weak fields as follows:
\begin{align}
	\partial_z A_n(z)&=\bigg(-\alpha + i\gamma \big(|A_n(z)|^2+\sum_{p\ne n}2|A_p(z)|^2\big)\bigg)A_n(z),\\
	\partial_z b_n(z)&=\bigg(-\alpha +2i\gamma\sum_p |A_p(z)|^2\bigg)b_n(z)+2i\gamma\sum_{l\ne n}A_l(z)A_n^*(z)b_l(z).
\end{align}

Using the same technique as Eq. (\ref{eqn:intensity_dynamics}), it follows that $|A_n(z)|=e^{-\alpha z}|A_n(0)|$,
and reinserting this with $A_n(0)=\sqrt{P}e^{i\theta_n}$ one obtains a loss-dependent nonlinear phase for the pumps:
\begin{equation}
	A_n(z)=\sqrt{P}e^{i\theta_n}e^{-\alpha z+i\phi_\alpha(z)(N-1/2)},
\end{equation}
where:
\begin{equation}
	\phi_\alpha(z)= 2\gamma P z e^{-\alpha z}\text{sinhc}(\alpha z).
\end{equation}
The dynamics of the weak fields then become:
\begin{align}
	\partial_z b_n(z)&=\bigg(-\alpha +2i\gamma N P e^{-2\alpha z}\bigg)b_n(z)+2i\gamma P e^{-2\alpha z}\sum_{l\ne n}e^{i\theta_l-i\theta_n}b_l(z).
\end{align}
In the frame $\tilde{b}_n(z):=e^{\alpha z+i\theta_n}b_n(z)$, this simplifies to:
\begin{align}
	\partial_z \tilde{b}_n(z)&=2i\gamma N P e^{-2\alpha z}\tilde{b}_n(z)+2i\gamma P e^{-2\alpha z}\sum_{l\ne n}\tilde{b}_l(z).
\end{align}

\noindent This is a matrix equation of the form $\partial_z\vec{\tilde{b}}(z)=iH(z)\vec{\tilde{b}}(z)$. Since $[H(z),H(z')]=0$, we may directly integrate:
\begin{align}
	\vec{\tilde{b}}(z)=e^{i\int_0^z dz'\,H(z)}\vec{\tilde{b}}(0).
\end{align}
Repeating the above procedures, the solution is:
\begin{equation}
    b_n(z)=\sum_l e^{-\alpha z + i\left[-\theta_n+\theta_l+i\phi_\alpha(z)(N-1)\right]}\left(\frac{e^{i\phi_\alpha(z)}-1}{N}+\delta_{nl}\right)b_l(0).
\end{equation}
Note that, for a single-mode input in mode $k$, the intensity is given by:
\begin{equation}
    |b_n(z)|^2=e^{-2\alpha z}\left|\tfrac{e^{i\phi_\alpha(z)}-1}{N}+\delta_{nk}\right|^2.
\end{equation}
In this experiment, the propagation length is fixed at $z=L$; only the initial pump power $P$ is varied. The resulting singles counts (measuring relative intensities versus pump power) for each channel are therefore scaled by one, \textit{constant} factor of $e^{2\alpha z}$ with a scaled nonlinear phase ($\phi_\alpha(z)$ instead of $\phi(z)$). That is, the intensity curves are identical to the lossless solutions up to a vertical and horizontal scale. In this experiment, propagation losses are quoted from the Vistacor datasheet provided by Corning: having at most $0.43$ dB/km at 1285 nm, the 100-m fiber has a total loss of $e^{-2\alpha L}\approx 1\%$. This justifies the use of the lossless theory curves to fit the intensities (described later in further detail), with the understanding that any corrections under the above assumptions merely augment the vertical and horizontal scales minimally (about 1\% and 0.5\%, respectively).

\subsection{Case of $N=3$}

In our experiment we study the case of Bragg scattering with three pumps. We set the pump powers to be equal and choose frequencies such that $\Delta k_2=0$ and $\Delta k_1$ is negligible compared to $\pi/L$, where $L$ is the interaction length (100 m in this experiment). Setting $N=3$ we have the solution:

\begingroup
\renewcommand{\arraystretch}{1.5}
\begin{equation}
    \begin{bmatrix}
        b_1(L)e^{i\theta_1}\\
        b_2(L)e^{i\theta_2}\\
        b_3(L)e^{i\theta_3}
    \end{bmatrix}
    =
    e^{2i\phi}
    \begin{bmatrix}
        p(\phi) & q(\phi) & q(\phi)\\
        q(\phi) & p(\phi) & q(\phi)\\
        q(\phi) & q(\phi) & p(\phi)
    \end{bmatrix}
    \begin{bmatrix}
        b_1(0)e^{i\theta_1}\\
        b_2(0)e^{i\theta_2}\\
        b_3(0)e^{i\theta_3}
    \end{bmatrix},
    \label{eqn:idealBS}
\end{equation}
\endgroup

\noindent where, as above, $\phi=2\gamma L P$ is the nonlinear phase and
\begin{align}
    q(\phi)&=\frac{e^{3i\phi}-1}{3}\\
    p(\phi)&=q(\phi)+1.
\end{align}

\subsection{Bragg-Scattering for Quantized Fields}
The quantized displacement field may be written (for instance, see \cite{quesada:22}):

\begin{equation}
    \hat{D}(\vec{r},t)=\int d\beta \,  \sqrt{\frac{\hbar \omega(\beta)}{4\pi}}\at(\beta,t)d(x,y)e^{i\beta z} + h.c.,
\end{equation}

\noindent where the field is approximated as exciting only the fundamental linearly polarized mode (e.g. $\text{HE}_{01}$) of the nonlinear fiber, with longitudinal coordinate $z$. In the above definition $\at(\beta,t)$ annihilates a plane wave with propagation constant $\beta$ at time $t$, having the equal-time commutator:
\begin{equation}
    [\at(\beta,t),\at^{\dagger}(\beta',t)]=\delta(\beta-\beta').
\end{equation}

Using the definition of the slowly-varying envelope in terms of which the NLSE is written \cite{agrawal2019},
\begin{equation}
    E(\Vec{r},t)=\tfrac{1}{2}F(x,y)A(z,t)e^{i(\beta_0 z-\omega_0 t)}+c.c.,
\end{equation}
and the respective spatial mode normalizations of \cite{quesada:22} and \cite{agrawal2019},
\begin{equation}
    \int dxdy \,\frac{|d(x,y)|^2}{\epsilon_0\epsilon(\omega)}\frac{v_p(\omega)}{v_g(\omega)}=1=\int dxdy\,\frac{|F(x,y)|^2}{\mathcal{A}},
\end{equation}
\noindent (where $v_p$, $v_g$ are phase and group velocities and $\mathcal{A}$ is the effective mode area), we identify the quantized slowly-varying envelope operator:
\begin{equation}
    \hat{A}(z,t)=\int d\beta\,\sqrt{\frac{\hbar\omega(\beta) v_g}{\pi\mathcal{A}\epsilon_0\epsilon(\beta)v_p}}\at(\beta,t)e^{i(\beta -\beta_0)z+i\omega_0 t}.
    \label{eqn:envlp_op}
\end{equation}
With the definition Eq. (\ref{eqn:Adef}) we may decompose the slowly-varying envelope into bands corresponding to the longitudinal envelope annihilation operator of the $i^{th}$ weak field:
\begin{equation}
    \hat{b}_i(z,t)=\int\displaylimits_{\beta_i-\Delta\beta/2}^{\beta_i+\Delta\beta/2} d\beta\,\sqrt{\frac{\hbar\omega(\beta) v_g}{\pi\mathcal{A}\epsilon_0\epsilon(\beta)v_p }} \at(\beta,t)e^{i(\beta-\beta_i)z+i\omega_i t},
    \label{eqn:wkfld_op}
\end{equation}

\noindent in correspondence with the classical fields $b_i$ in which the Bragg-scattering transformation Eq. (\ref{eqn:weakfielddynamics}) is written (note the relative factor of $e^{i\delta\beta_i z-i\delta\omega_i t}$ between Eqs. (\ref{eqn:wkfld_op}) and (\ref{eqn:envlp_op})). Neglecting the frequency dependence of the root in the integrand of Eq. (\ref{eqn:wkfld_op}), we rewrite the longitudinal envelope operator as
\begin{equation}
    \hat{b}_i(z,t)\approx\sqrt{C}\int\displaylimits_{-\Delta\beta/2}^{\Delta\beta/2} \frac{d\beta}{\sqrt{2\pi}}\,\at_i(\beta,0)e^{i\beta z-i(\omega(\beta_i+\beta)-\omega_i))t},
\end{equation}
where
\begin{align}
    \at_i(\beta,t)&:=\at(\beta_i+\beta,t),\\
    \text{and}\quad C&:=\frac{2\hbar\omega_0 v_{g,0}}{\mathcal{A}\epsilon_0\epsilon(\omega_0)v_{p,0} }.
\end{align}

\noindent The free (i.e. non-interacting) spatial annihilation operators may then be defined as:
\begin{align}
    \ha_i(z,t):=\int\displaylimits_{-\Delta\beta/2}^{\Delta\beta/2} \frac{d\beta}{\sqrt{2\pi}}\,\at_i(\beta,0)e^{i\beta z-i(\omega(\beta_i+\beta)-\omega_i))t}\\
    [\ha_i(z,t),\ha_j^{\dagger}(z',t')]\approx\delta_{ij}\delta(z-z'-v_{g,i}(t-t')).
\end{align}

Because the classical equation of motion Eq. (\ref{eqn:Npump}) is linear in the weak fields, the classical and quantum dynamics are identical. We may therefore replace the classical interacting fields with bosonic annihilation operators in the Heisenberg picture. \textit{In the interaction region} (i.e. the nonlinear medium), under the assumptions of negligible phase mismatch and equal pump powers, the position operators at propagation distance $z=L$ and $z=0$ are related by\footnote{(Taking the group velocities as approximately constant, $v_{g,i}\approx v_g$.)}:
\begin{equation}
    \ha_i(L,t)=\sum_j U_{ij}(\phi)\ha_j(0,t-L/v_g),
\end{equation}
\noindent where $U_{ij}(\phi)$ is given from Eq. (\ref{eqn:idealBS}) as
\begin{equation}
    U_{ij}(\phi)=
    \begin{cases}
        p(\phi)&\quad i=j\\
        q(\phi)e^{i(\theta_j-\theta_i)}&\quad i\ne j.
    \end{cases}
    \label{eqn:3pbs}
\end{equation}

\noindent The phase factor $e^{2i\phi}$ from Eq. (\ref{eqn:idealBS}) is a global phase at a given propagation length $L$ and may be absorbed into the definition of $\ha_i(L,t)$.

\section{Correlation Functions}

The spectrally filtered field measured at one of the detectors corresponds to the operator $\ha_i(z,t)$, where $i\in\{1,2,3\}$ denotes one of the three interacting frequency modes. The singles rates are then proportional to the first-order correlation function at the detectors given by the expectation:

\begin{equation}
    G_i^{(1)}(z)=\braket{\psi|\ha_i^{\dagger}(z,t)\ha_i(z,t)|\psi}.
\end{equation}

\noindent Coincidence rates are proportional to the second-order correlation function:

\begin{equation}
    G_{ij}^{(2)}(z,\tau)=\braket{\psi|\ha_j^{\dagger}(z,t)\ha_i^{\dagger}(z,t+\tau)\ha_i(z,t+\tau)\ha_j(z,t)|\psi}.
\end{equation}

\noindent To compute the total coincidence count rate we integrate the coincidence histogram over the inverse linewidth of the photon pair source (with respect to the event delay time, $\tau$). Since the integration washes out the temporal dependence of the fields and absorbs the Dirac-delta functions in the field commutators, for simplicity we adopt a discrete-mode model by using the following replacement rules:

\vspace{5 mm}
\begin{tabular}{rcl}
    $\ha_i(z,t)$&$\rightarrow$&$\ha_i(z)$\\
    $[\ha_i(z,t),\ha_j^{\dagger}(z,t')]=\delta_{ij}\delta(t-t')/v_{g,i}$&$\rightarrow$ & $[\ha_i(z),\ha_j^{\dagger}(z)]=\delta_{ij}$
\end{tabular}
\vspace{5 mm}

 \noindent Thus, the integrated first and second-order correlation functions may be computed using:
 \begin{align}
    G_i^{(1)}(z)&=\braket{\psi|\ha_i^{\dagger}(z)\ha_i(z)|\psi}=||\ha_i(z)\ket{\psi}||^2,\\
    G_{ij}^{(2)}(z)&=\braket{\psi|\ha_j^{\dagger}(z)\ha_i^{\dagger}(z)\ha_i(z)\ha_j(z)|\psi}=||\ha_i(z)\ha_j(z)\ket{\psi}||^2.
\end{align}
 \noindent The discrete-mode model contains the essential physics necessary to compute the coincidence rates as a function of nonlinear phase. In order to investigate finer temporal dependence (e.g. to compute the dependence of the correlation function on delay time $\tau$), the time dependence would need to be retained.

\subsection{Classical Input}

\subsubsection{Single-Mode Coherent State}

For the coherent state $\ket{\nu_1}= e^{\nu_1 \ha_1^{\dagger}(z=0)-h.c.}\ket{0}$ occupying the first frequency mode as input, the singles rates at $z=L$ are given by:

\begin{equation}
    G_i^{(1)}(z)=\braket{\nu_1|\ha_i^{\dagger}(L)\ha_i(L)|\nu_1}.
\end{equation}

\noindent Using the linear transformation $\ha_i(z)=\sum_k U_{ik}(z)\ha_k(0)$ where $\ha_i(0)$ denotes the $i^{th}$ field before interaction, we may write

\begin{equation}
    G_i^{(1)}(z)=||\sum_j U_{ij}(z)\ha_j(0)\ket{\nu_1}||^2=|U_{i1}(z)|^2|\nu_1|^2.
\end{equation}
\noindent Explicitly, in terms of the nonlinear phase $\phi=2\gamma L P$:
\begin{equation}
\begin{split}
    G_1^{(1)}(\phi)&=|\nu_1|^2|U_{11}(\phi)|^2=|\nu_1|^2|p(\phi)|^2,\\
    G_2^{(1)}(\phi)&=G_3^{(1)}(\phi)=|\nu_1|^2|U_{21}(\phi)|^2=|\nu_1|^2|q(\phi)|^2.
\end{split}
\end{equation}

\subsubsection{Dual-Mode Coherent State}

For a dual-mode CW coherent state (omitting the $z$ argument, $\ha:=\ha(z=0$))
\begin{equation}
    \ket{\nu_1,0,\nu_3}=e^{\nu_1 \ha^{\dagger}_1+\nu_3 \ha^{\dagger}_3-h.c.}\ket{0},
\end{equation}
\noindent the first-order correlation function following the interaction is
\begin{equation}
\begin{split}
    G_i^{(1)}(z)&=||\sum_j U_{ij}(z)\ha_j\ket{\nu_1,0,\nu_3}||^2\\
    &=|U_{i1}(z)\nu_1+U_{i3}(z)\nu_3|^2.
\end{split}
\end{equation}

Since in our experiment the relative phase between the two coherent states is not fixed, we write $\nu_3=\nu_1 e^{i\vartheta(t-\beta_1 z)}$. Final results must then be averaged over the independent classical random variables $\vartheta(t-\beta_1 z)$ (i.e. the relative phases $\{\vartheta_k\}$ during the $k^{th}$ interaction window), uniformly distributed over $[0,2\pi)$. Since the pump phases $\theta_i$ contained in $U_{ij}$ are also randomly fluctuating, these too must be averaged over. Finally, in our experiment we choose $|\nu_1|=|\nu_3|=\nu$. Using the above formalism, after averaging we obtain:
\begin{equation}
\begin{split}
    G_{1}^{(1)}(z)&=G_{3}^{(1)}(z)=\nu^2\left(|p(z)|^2+|q(z)|^2\right)\\
    &\hspace{16 mm}=\nu^2(1-|q(z)|^2)\\
    G_{2}^{(1)}(z)&=2\nu^2|q(z)|^2.
\end{split}
\end{equation}

\noindent The total coincidence rate is\footnote{Note that, before integrating out the explicit time dependence, the coinciding events at detectors $i$ and $j$ occur at some delay $\tau$; therefore the relative phase variables $\{\vartheta(t),\vartheta(t+\tau)\}$ are independent when averaged over all interaction windows.}:
\begin{equation}
\begin{split}
    G_{ij}^{(2)}(z)&=||\ha_i(z)\ha_j(z)\ket{\nu_1,0,\nu_3}||^2\\
    &=|(U_{i1}(z)\nu_1+U_{i3}\nu_3(z))(U_{j1}(z)\nu_1+U_{j3}(z)\nu_3)|^2\\
    &=G_i^{(1)}(z)G_j^{(1)}(z).
\end{split}
\end{equation}

\noindent We define a normalized coincidence rate as:
\begin{equation}
    g^{(2)}_{ij}(\phi):=\frac{G_{ij}^{(2)}(\phi)}{G_{ij}^{(2)}(0)}.
    \label{eqn:norm_coinc_def}
\end{equation}

\noindent The normalized coincidence for ports one and three is then:
\begin{equation}
    g^{(2)}_{13}(\phi)=\left(1-|q(\phi)|^2\right)^2.
\end{equation}

\subsection{Photon-Pair Input}

Consider now the two-photon state as input:
\begin{equation}
    \ket{\psi(0)}=\ha^{\dagger}_1(0)\ha^{\dagger}_3(0)\ket{0}.
\end{equation}
The singles rates are given by:
\begin{equation}
    G^{(1)}_i(z)=|U_{i1}(z)|^2+|U_{i3}(z)|^2.
\end{equation}

\noindent Note that this expression is identical to that of two coherent states that are mutually phase-incoherent.

The coincidence rates, however, are given by:
\begin{equation}
    G_{ij}^{(2)}(\phi)=|U_{i1}(\phi)U_{j3}(\phi)+U_{i3}(\phi)U_{j1}(\phi)|^2.
\end{equation}

\noindent The coherent sum appearing in this expression reflects the indistinguishability of the two photons; the two terms account for the amplitudes that the photons exiting at the $i^{th}$ and $j^{th}$ ports originated respectively from input ports 1 and 3 or vice versa.

The normalized coincidence rate is then:
\begin{align}
    g^{(2)}_{13}(\phi)&=\left|p^2(\phi)+q^2(\phi)\right|^2,
    \label{eqn:2photon_g13}
    \\
    g^{(2)}_{12}(\phi)&=g^{(2)}_{23}(\phi)=|p(\phi)q(\phi)+q^2(\phi)|^2.
\end{align}

\subsection{Including Multiphoton Effects and Loss}

Using the simplified discrete-mode model it is straightforward to incorporate the multiphoton contribution inherent in photon-pair generation as well as system insertion losses.

The spontaneous four-wave mixing process used in our experiment to generate photon pairs can be modeled using a two-mode squeezed vacuum state $\ket{\zeta}=\hat{U}(\zeta)\ket{0}$ where
\begin{equation}
    \hat{U}(\zeta)=e^{\zeta \ha_1^{\dagger}(0)\ha_3^{\dagger}(0)-h.c.}.
\end{equation}
\noindent Evolving the fields as $\ha_i(\zeta):=\hat{U}^{\dagger}(\zeta)\ha_i(0)\hat{U}(\zeta)$, the squeezing operator $\hat{U}$ leads to the following transformation in the Heisenberg picture:
\begin{equation}
    \begin{bmatrix}
        \ha_1(\zeta)\\
        \ha_3^{\dagger}(\zeta)
    \end{bmatrix}
    =
    \begin{bmatrix}
        \cosh(|\zeta|) & \frac{\zeta}{|\zeta|}\sinh(|\zeta|) \\
        \frac{\zeta^*}{|\zeta|}\sinh(|\zeta|) & \cosh(|\zeta|)
    \end{bmatrix}
    \begin{bmatrix}
        \ha_1(0)\\
        \ha_3^{\dagger}(0)
    \end{bmatrix}.
\end{equation}

We include loss using a beamsplitter model which couples a given fiber mode $\ha_i$ to its own external vacuum mode with annihilator $\hat{r}_i$:
\begin{equation}
    \hat{U}(\alpha)=e^{\Sigma_{i}\alpha_i\ha_i^{\dagger}\hat{r}_i-h.c.}
\end{equation}

\noindent The loss may be regarded as equivalent to a non-unit transmission coefficient, $T_i=\cos(|\alpha_i|)$, corresponding to a reflection coefficient $R_i=\frac{\alpha_i}{|\alpha_i|}\sin(|\alpha_i|)$. The Heisenberg-picture transformation is then given by:
\begin{equation}
    \begin{bmatrix}
        \ha_i(\alpha)\\
        \hat{r}_i(\alpha)
    \end{bmatrix}
    =
    \begin{bmatrix}
        T_i & R_i \\
        -R_i^* & T_i
    \end{bmatrix}
    \begin{bmatrix}
        \ha_i(0) \\
        \hat{r}_i(0)
    \end{bmatrix}.
\end{equation}

\noindent We account for losses before and after the Bragg-scattering interaction using the parameters $\alpha_i$ and $\mu_i$ respectively.

The full evolution of the state prior to detection is then given by
\begin{equation}
    \ket{\psi}_{final}=\hat{U}(\mu)\hat{U}(\phi)\hat{U}(\alpha)\hat{U}(\zeta)\ket{0},
\end{equation}
\noindent where $\zeta$ and $\phi$ parametrize the squeezing and Bragg-scattering evolution, respectively.

The singles rates are given by
\begin{equation}
    \begin{split}
        G^{(1)}_i(\phi)&=||\ha_i\hat{U}(\mu)\hat{U}(\phi)\hat{U}(\alpha)\hat{U}(\zeta)\ket{0}||^2\\
        &=|T_i(\mu)|^2\sinh^2(|\zeta|)\sum_{j\ne2} |U_{ij}(\phi)T_j(\alpha)|^2.
    \end{split}
\end{equation}

The coincidences are
given by

\begin{equation}
    \begin{split}
        G^{(2)}_{ij}&=||\ha_i\ha_j\hat{U}(\mu)\hat{U}(\phi)\hat{U}(\alpha)\hat{U}(\zeta)\ket{0}||^2\\
        &=|T_i(\mu)|^2|T_j(\mu)|^2|T_1(\alpha)|^2|T_3(\alpha)|^2\\
        &\hspace{0.5 cm}\times\bigg\{\bigg|U_{i,1}(\phi)U_{j,3}(\phi)+U_{i,3}(\phi)U_{j,1}(\phi)\bigg|^2\bigg(\sinh^2(|\zeta|)+2\sinh^4(|\zeta|)\bigg)\\
        &\hspace{-0.5 cm}+2\bigg(|U_{i,1}(\phi)|^2|U_{j,1}(\phi)|^2\frac{|T_1(\alpha)|^2}{|T_3(\alpha)|^2}+|U_{i,3}(\phi)|^2|U_{j,3}(\phi)|^2\frac{|T_3(\alpha)|^2}{|T_1(\alpha)|^2}\bigg)\sinh^4(|\zeta|)\bigg\}.
    \end{split}
\end{equation}

Inserting the expressions for $U_{ij}(\phi)$ from Eq. (\ref{eqn:3pbs}), we obtain the normalized coincidence rate (considering for simplicity only coincidences between output frequency modes one and three):
\begin{equation}
    \begin{split}
        g^{(2)}_{13}(\phi)&=|p^2(\phi)+q^2(\phi)|^2\\
        &\hspace{1 cm}+2|p(\phi)|^2|q(\phi)|^2\left(\frac{|T_1(\alpha)|^2}{|T_3(\alpha)|^2}+\bigg|\frac{T_3(\alpha)}{T_1(\alpha)}\bigg|^2\right)\left(\frac{\sinh^2(|\zeta|)}{1+2\sinh^2(|\zeta|)}\right).
    \end{split}
    \label{eqn:g13_multiphoton}
\end{equation}

\noindent By taking the photon-pair limit of this expression, one obtains the two-photon prediction of Eq. (\ref{eqn:2photon_g13}). The first term in the RHS therefore corresponds to photon-pair statistics, and the second may be understood as originating from multiphoton effects. Note that appearance of asymmetry in the input losses $T_{i}(\alpha)$ only enters via the multiphoton term. This is physically intuitive since, in the photon-pair limit, when one photon in the pair is lost the coincidence is not counted; so normalized coincidence rates do not depend on loss asymmetry. With multiphoton effects, however, even when one photon is lost coincidences may still be registered.

\begin{figure}[ht]
\centering
\includegraphics[width=0.8\textwidth]{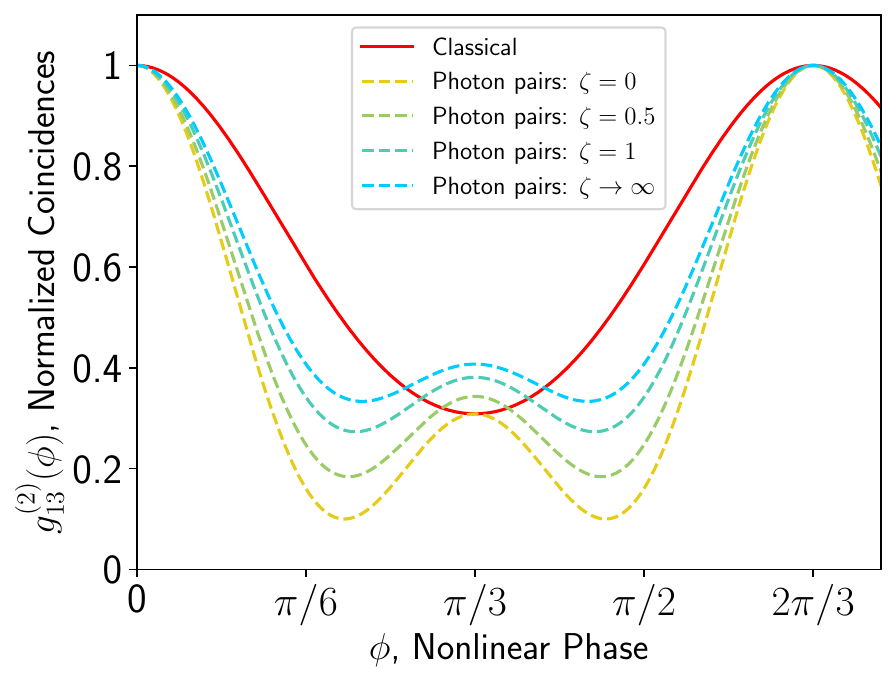}
\caption{Theoretical coincidence rates for two classical (coherent-state) inputs and a photon-pair input interfering via the three-way Bragg scattering interaction. $\zeta$ represents the multiphoton parameter.}
\label{SM_theory_curves}
\end{figure}

The resulting theoretical normalized coincidences of the classical two-mode input (phase-averaged coherent states) and photon-pair input (including multiphoton contribution with symmetric losses) are plotted in Fig. \ref{SM_theory_curves}. Whereas the classical curves are given by the products of the two uncorrelated singles rates, for photon pairs an interference effect is predicted. This contrast in coincidence rates comes about due to the interference of amplitudes corresponding to interchangeable two-photon states, and may be interpreted as a Hong-Ou-Mandel effect.

\section{Measurement of Model Parameters}

\subsection{Losses Before Nonlinear Fiber}

According to Eq. (\ref{eqn:g13_multiphoton}), to account for the contributions due to multiphoton statistics it is necessary to incorporate the ratio of losses accumulated between generation and before Bragg scattering. We estimate these losses using measurements concurrent with the experiment. The dominant sources of loss originate from wavelength-dependent asymmetries in the filtering setup prior to the nonlinear fiber. To measure the ratio of the losses before the fiber, we first measure the relative insertion losses of the filtering setup following the nonlinear fiber. We then factor this out of the singles rates measured during each trial. Assuming negligible differences in SNSPD detection efficiency, this results in an estimate of the loss ratio prior to entering the fiber. We can then use this measurement in our model Eq. (\ref{eqn:g13_multiphoton}), eliminating an otherwise free parameter.

\subsection{Multiphoton Parameter}

We generate photon pairs using spontaneous four-wave mixing by pumping a single resonance (at approximately 1282.8 nm) of an integrated SiN microresonator. Microscopically, two pump photons are annihilated and correlated photons with frequencies at $\pm\Delta\omega$ are created. However, photon pairs are only the leading-order effect; the full state contains multiphoton terms and is better described by the two-mode squeezed vacuum (see e.g. \cite{ou2017}):
\begin{equation}
\ket{\zeta}=e^{\zeta\ha^{\dagger}_+\ha^{\dagger}_--h.c.}\ket{0}=\text{sech}(|\zeta|)\sum_{n=0}^{\infty}\left(\frac{\zeta}{|\zeta|}\tanh(|\zeta|)\right)^n\ket{n,n}.
\end{equation}

\begin{figure}[!hb]%
\centering
\includegraphics[width=0.5\textwidth]{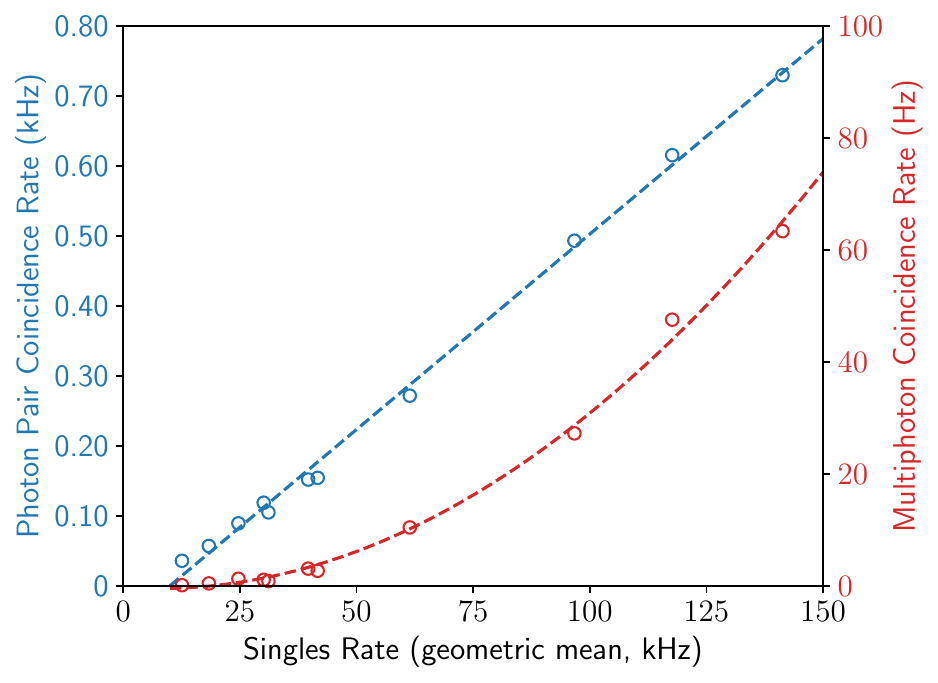}
\caption{Scaling of photon-pair and multiphoton coincidences with singles rate. Dashed lines indicate linear and quadratic fits to the data (circles).}
\label{multiphoton_scaling}
\end{figure}

\begin{figure}[ht]%
\centering
\includegraphics[width=0.6\textwidth]{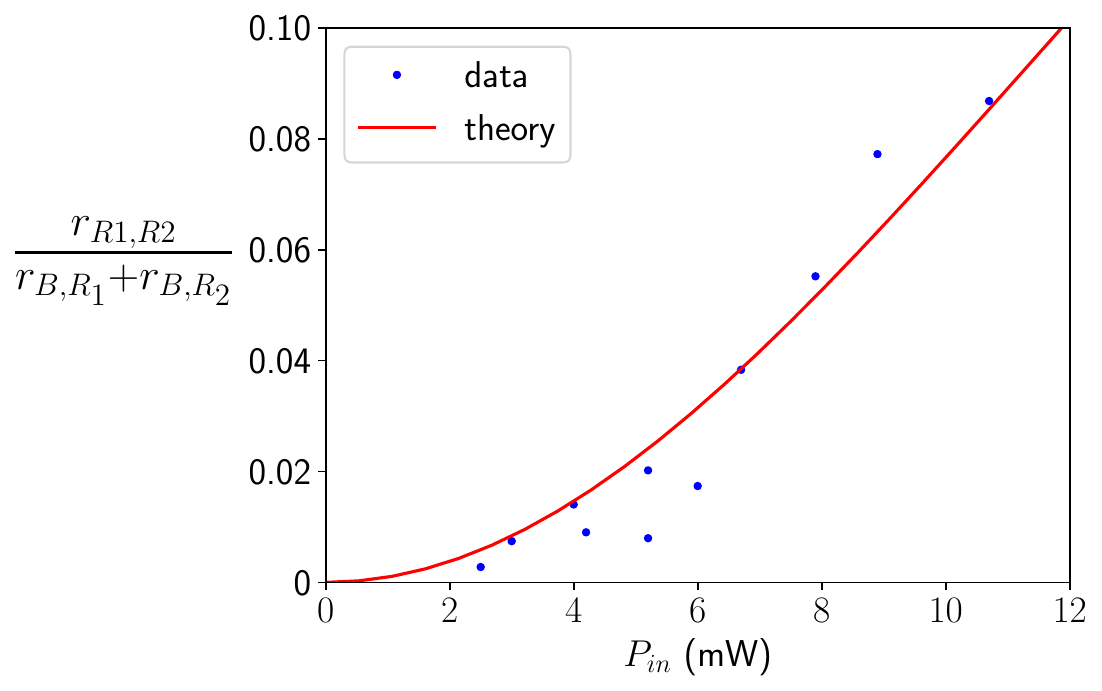}
\caption{Ratio of multiphoton coincidences (B1-B2) to dual-frequency coincidences (A-B1 plus A-B2).}
\label{multiphoton}
\end{figure}

To estimate $|\zeta|$, we measure multiphoton coincidences as a function of pump power. The output of the ring is first spectrally filtered into two channels, R (red photon, 1285.0 nm) and B (blue photon, 1280.6 nm). (The freespace filtering setup after the nonlinear fiber is used rather than fiber-Bragg gratings; this helps avoid introducing the loss asymmetry described above.) Channel R is then sent through a 50:50 fiber beamsplitter resulting in two output channels, R1 and R2. If only pairs of single photons (i.e., $\ket{1,1}$) were present, then no coincidences should be registered between channels R1 and R2; only B-R1 and B-R2 coincidences would be detected. R1-R2 coincidences therefore serve to characterize the multiphoton emission. In Fig. \ref{multiphoton_scaling} are plotted multiphoton ($r_{R1,R2}$) and photon pair ($r_{B,R1}+r_{B,R2}$) coincidence rates versus singles rates ($\sqrt{r_B(r_{R1}+r_{R2})}$). The expected scaling is observed: photon-pair coincidences scale linearly with the singles rate, $r_{pair}\propto r_{sgls}\propto|\zeta|^2$, whereas multiphoton coincidences are approximately quadratic, $r_{mult}\propto r_{sgls}^2\propto|\zeta|^4$. Corresponding linear and quadratic fits are plotted in agreement with theory.

SFWM pump power is varied from 2.5 to 10.7 mW (measured before the chip). The experiment is performed at 10.8 mW of pump power. Multiphoton emission is characterized by plotting the ratio of R1-R2 coincidences to the sum of B-R1 and B-R2 coincidences. The results are shown in Fig. \ref{multiphoton}. From fitting to the analytical curve we obtain $|\zeta|\approx0.4$.

\section{Data Analysis}
\subsection{Single-Frequency Input}

In order to compare experiment with theory, we perform least-squares fits of detection events as a function of peak pump power to the sinusoidal curves obtained analytically under the assumptions of equal pump powers, zero phase mismatch ($\Delta k_n = 0$), and lossless nonlinear fiber.

\begin{figure}[ht]
\centering
\includegraphics[width=\textwidth]{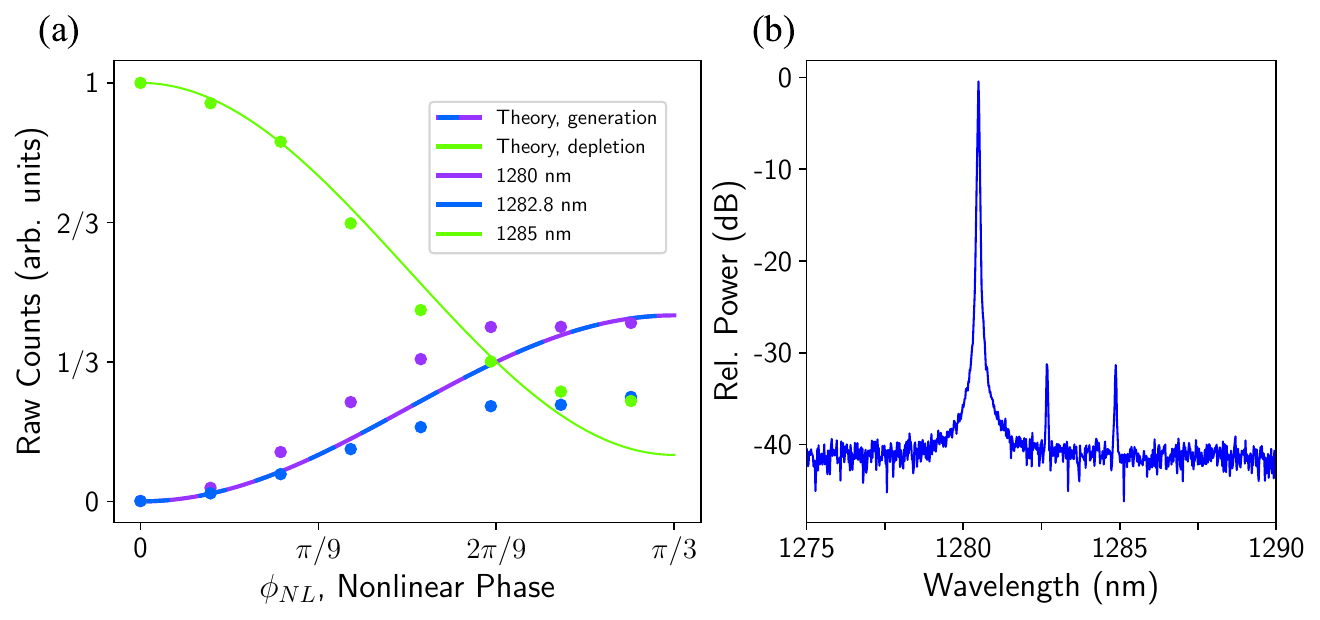}
\caption{Raw singles rates as a function of power, normalized to input signal rate at zero pump power.}
\label{sgls_raw}
\end{figure}

The raw singles rates obtained for classical single-frequency input in Fig. \ref{sgls_raw} are for each data point normalized to the input signal rate at zero pump power (i.e. without Bragg scattering). It is necessary to convert from measured peak powers of the Bragg pumps to nonlinear phase in order to fix the horizontal scale.

Since the depletion curve (the rates measured for the input frequency as it is converted to the other two wavelengths) can be normalized to 1 at zero pump power, only one free parameter is needed to convert the horizontal scale from the pump peak power to the nonlinear phase. To determine the horizontal scale, we fit the normalized depletion curve to the analytical curve (under the above ideal assumptions). The peak pump power of the three pumps, measured directly from oscilloscope traces for each data point, can then be converted to nonlinear phase. This in turn fixes the nonlinear phase for all three curves.

As the two generation curves start at zero counts at zero nonlinear phase, they cannot be normalized in the same way as the depletion curve without assuming the absolute rate at some nonzero nonlinear phase. This would entail an impractical measurement of the individual losses for each channel following the nonlinear fiber, which is unreliable due to the dependence of the freespace filters and detectors on polarization. The polarization changes frequently as it is optimized during each run of the experiment. Rather than attempt to correct for these losses directly, we use the fact that the loss is fixed for a given channel throughout the experiment, meaning that there is a single loss factor for each depletion curve. To obtain this factor, we perform a least-squares fit to the corresponding theory curves using the previously determined nonlinear phase, resulting in Fig. 3 of the main text. Note that, as derived earlier, the presence of propagation loss changes only the vertical and horizontal scale but not the shape of the curves; furthermore, using the propagation loss of 0.43 dB/km at 1285 nm provided by Corning for our Vistacor fiber, these scales deviate from the true scales by $<1\%$.

In the classical regime (where the mean photon number is 10s of decibels higher than the latching limit of the SNSPDs), we readily verify that the two generation curves are in fact symmetric by sending the output of the nonlinear fiber directly into an optical spectrum analyzer (OSA). While filtering the individual channels inevitably introduces asymmetric losses, the OSA can measure a full spectrum with uniform insertion losses. The spectrum in Fig. 3 of the main text strongly verifies the symmetric distribution of power in the three channels after accounting for the duty cycle of the input signal. Further deviations from theory in Fig. 3 of main text are mostly due to the departure from the assumptions of zero phase mismatch and equal pump powers.

\subsection{Dual-Frequency Input}
To generate the coincidence curves of Fig. 4 (main text), for both classical and quantum coincidence measurements we follow a similar procedure as for the single-frequency input case. The raw coincidence curves are first normalized by the concurrent accidental rates at zero power, that is, the product of the corresponding singles rates taken at a time window at fixed relative delay with respect to the pump pulses such that the pumps are off. This quantity may be written
\begin{equation}
    R_{coinc}=\frac{\braket{\ha^\dagger_3\ha^\dagger_1\ha_1\ha_3}}{\braket{\ha^\dagger_1\ha_1}_0\braket{\ha^\dagger_3\ha_3}_0},
\end{equation}

\noindent where the 0-subscript in $\braket{\cdot}_0$ denotes the quantity in the absence of Bragg scattering, at zero pump power. This quantity is proportional to the integrated second-order correlation function. To complete the normalization such that the coincidence curve is equal to 1 at zero nonlinear phase, for all data points we normalize with respect to this same quantity at zero nonlinear phase. Finally, the conversion factor from relative pump pulse power to nonlinear phase is set by fitting the normalized, respective \textit{singles} curves to the analytical curves, using the independently-measured pre-fiber losses and multiphoton parameters for the case of photon-pair input.

\bibliography{supplementary-bibliography}